\documentstyle[12pt]{article}
\begin{document}
\voffset -2.5cm
\hoffset -1cm
\baselineskip 24pt
\newcommand\be{\begin{equation}}
\newcommand\ee{\end{equation}}
\title{Canonical Phase Diagrams of the 1-D Falicov-Kimball Model
at T=0}
\author{
\sl Z.\ Gajek\\
Institute of Low Temperature \\  and Structure Research of
Polish Academy of Sciences,\\
P.\ O. Box 937, 50--950 Wroc{\l}aw, Poland\\
\and
\sl J.\ J\c{e}drzejewski  \\
Institute of Theoretical Physics, \\
University of Wroc{\l}aw,\\
Max Born Sq.\ 9, 50--204 Wroc{\l}aw, Poland
\and
\sl  R.\ Lema\'{n}ski$^{*}$ \\
Institute of Low Temperature \\  and Structure Research of
Polish Academy of Sciences,\\
P.\ O. Box 937, 50--950 Wroc{\l}aw, Poland\\}
\date{ }
\maketitle
\begin{abstract}

The Falicov-Kimball model
of spinless quantum electrons hopping on a 1-di\-men\-sio\-nal
lattice and  of immobile classical ions occupying some lattice
sites, with only intrasite coupling between those particles,
have been studied at zero tem\-pe\-ra\-tu\-re
by means of well-controlled numerical procedures.
For selected values of the unique coupling parameter $U$ the
restricted phase dia\-grams (based on all the periodic configurations
of localized particles (ions) with period not greater than
16 lattice constants, typically) have been  constructed in the
grand-ca\-no\-ni\-cal ensemble. Then these dia\-grams have been
translated into the ca\-no\-ni\-cal ensemble.
Compared to the diagrams obtained in other studies our ones
contain more details, in particular they give better insight
into the way the mixtures of periodic phases are formed.
Our study has revealed several families of new characteristic
phases like the generalized most homogeneous and the generalized
crenel phases, a first example of a structural phase transition
and a tendency to build up an  additional symmetry -- the
hole-particle symmetry with respect to the ions (electrons) only,
as $U$ decreases.

\end{abstract}
\vspace{3mm}
$^{*}$formerly known as R.\  {\L}y\.{z}wa

\newpage

%***********************************************************************

\section{The model and the method}

The work reported in this paper has been inspired
by the recent results due to
Freericks and Falicov \cite{frefal},
Farka\v{s}ovsk\'{y} and Bat'ko \cite{farbat}, and
Gruber et al. \cite{guj}.
It is also a continuation of our earlier work
\cite{llj}. All the papers mentioned are
concerned with various properties of the ground state phase
diagram of the two-state spinless Falicov-Kimball model.
This is a lattice model of itinerant quantum electrons and
immobile classical ions that are coupled via a simple
on site interaction. It is a simplified version of the
model that was put forward in 1969 by Falicov and Kimball
\cite{falkim} to describe  conductor-insulator transitions.
In the one-dimensional case, considered in this paper,
the Hamiltonian corresponding to the model has the form
\begin{equation}
\label{ham}
H_{\Lambda}=\displaystyle{ \sum\limits_{x\in \Lambda} \left(
-a^{*}_{x} a_{x+1} - a^{*}_{x+1} a_{x} - U w(x) a^{*}_{x} a_{x}
\right),}
\end{equation}
where $\Lambda$ is a finite piece of the one-dimensional lattice
with periodic boundary conditions imposed,
$a^{*}_{x}, a_{x}$ are the creation and annihilation operators
of itinerant spinless electrons and $w(x)=0$ or $1$ is the
occupation number of the ions at the lattice site $x$.
The function $\Lambda \ni x \to w(x)$, is called the
{\em ion configuration}.
Observe that the electrons can hop only between
the nearest neighbour sites and the energy units have been chosen
in such a way that the hopping rate is unity.
Therefore in $H_{\Lambda}$ there is only one independent
parameter expressed in these units:
the electron-ion coupling $U$.
The total number of electrons
$N_{e}=\sum \limits_{x\in \Lambda} a^{*}_{x} a_{x}$ and
the local occupation numbers $w(x)$ of the ions
(hence the total number
$N_{i}(w)=\sum \limits_{x\in \Lambda} w(x)$ of the ions in the
configuration $w$) are the conserved quantities.

The Falicov--Kimball model can be related
to the celebrated Hubbard model \cite{kenlieb},
\cite{brandt}.
It has however many interpretations of its own
\cite{frefal}, \cite{kenlieb}, \cite{khom},
interesting enough to motivate intensive investigations
of this model.
In one of the interpretations proposed
by Khomskii \cite{khom}, Kennedy and Lieb \cite{kenlieb}
and adopted here it is a model of crystalline formation.
As proved
by Kennedy and Lieb \cite{kenlieb}
and by Brandt and Schmidt \cite{brasch}
the effective interaction between the ions,
induced by the direct interaction with mobile electrons,
leads to the formation of periodic structures of
the ions (superlattices).
Another interpretation that is sound from the point of view of
solid state physics, is in terms of mobile s--electrons
and localized $f$--electrons in rare--earth
compounds. This  makes the model relevant for
understanding mixed-valence phenomena \cite{falkim}, \cite{khom}.

For a given ion configuration $w$ the
Hamiltonian (\ref{ham}) is the second quantized form of the
{\em one-particle operator} $h(w)$ whose matrix elements are
\begin{equation}
\label{matrix_h}
[h(w)]_{xy}=-t_{xy} - U w(x) \delta_{xy},
\end{equation}
where
\begin{equation}
\label{matrix_t}
\begin{array}{l}
t_{xy} =
\cases{ 1 &\hbox{if} $y = x\pm 1$ \cr
0 &\hbox{otherwise}. }
\end{array}
\end{equation}
In the sequel by the {\em spectrum of the configuration $w$}
we mean the spectrum of $h(w)$.

The states of the whole system are characterized by their ion
configurations and the electrons' wave functions.
If the ion configuration $w$ and the total number of electrons
$N_{e}$ are specified, then in the ground state the electrons'
wave function corresponds to the so called {\em Fermi sea}.
The energy of this state, $E^{U}(w; N_{e})$,
called the {\em Fermi sea energy}, is equal
simply to the sum of all the $N_{e}$ lowest eigenvalues of $h(w)$.
Such states of the system are referred to as {\em phases}.
Suppose that we are given the total number $N_{e}$ of the electrons
and the total number $N_{i}$ of the ions.
In the ground state subject to these conditions the ions
are arranged in a specific configuration $\stackrel{\sim}{w}$,
the {\em ground state configuration}, and the electrons form the
corresponding Fermi sea.
This state of the whole system is referred to as the
$(N_{e},N_{i})$-phase.
Its energy, $E(N_{e},N_{i})$, called the {\em ground state energy},
is equal to $E^{U}(\stackrel{\sim}{w}; N_{e})$.
The largest eigenvalue of $h(\stackrel{\sim}{w})$ that contributes
to $E^{U}(\stackrel{\sim}{w}; N_{e})$, referred to as the
{\em Fermi level}, is separated from the higher and lower parts of
the spectrum of $\stackrel{\sim}{w}$ by the gaps
$\Delta_{u}(\stackrel{\sim}{w})$ and
$\Delta_{l}(\stackrel{\sim}{w})$, respectively.
A phase that is the  $(N_{e},N_{i})$-phase for some pair
$(N_{e},N_{i})$ is said to appear in the canonical phase diagram or
is referred to as {\em stable phase}.

Our results refer exclusively to the {\em infinite system}.
The thermodynamic limit is taken in such a way that
the electron number per site $N_{e}/| \Lambda |$,
the ion number per site $N_{i}/| \Lambda |$,
the Fermi sea energy per site  $E^{U}(w;N_{e})/| \Lambda |$
and the ground state energy per site $E(N_{e},N_{i})/|\Lambda |$
converge to the electron density $\rho_{e}$, the ion density
$\rho_{i}$, the Fermi sea energy density $e^{U}(w;\rho_{e})$
and the ground state energy density $e(\rho_{e},\rho_{i})$,
respectively.
Suppose that the ion configuration of the $(\rho_{e},\rho_{i})$-phase
is $\stackrel{\sim}{w}$. If in the thermodynamic limit
$\Delta_{u}(\stackrel{\sim}{w}) - \Delta_{l}(\stackrel{\sim}{w})>0$,
then the $(\rho_{e},\rho_{i})$-phase is called the
{\em insulator}, otherwise it is termed the {\em conductor}.

The possibility of studying directly infinite systems stems from
the fact that for the periodic ion configuration $w$
the Green function of the
tight binding Schr\"{o}dinger equation and consequently the density
of states $n(w)$, $\rho_{e}$ and $e^{U}(w;\rho_{e})$
can be determined exactly.
However the corresponding formulae depend on the
band edges of the spectrum of $w$, which are zeroes of the polynomial
whose order coincides with the period of $w$ \cite{lyz}.
First these zeros have to be determined numerically and after
that the quantities of interest,
like the energy density $e^{U}(w;\rho_{e})$, can be obtained by
a numerical integration.

The hole-particle unitary transformations applied to the
Hamiltonian (\ref{ham}) reveal the corresponding symmetries of
the Fermi sea energy density $e^{U}(w; \rho_{e})$.
Thus, performing the hole-particle transformation
with respect to the ions we obtain
\begin{equation}
\label{symm_i}
e^{(U)}\left(w^{*};\rho_{e}\right)
= e^{(-U)}\left(w;\rho_{e}\right) - U \rho_{e},
\end{equation}
where $w^{*}$ is the {\em adjoint} ion configuration of $w$,
$w^{*}(x)=1-w(x)$,
while with respect to the electrons we get
\begin{equation}
\label{symm_e}
e^{(U)}\left(w;\rho_{e}\right)
= e^{(-U)}\left(w; 1 - \rho_{e}\right) - U \rho_{i}(w).
\end{equation}
Therefore we can reduce the range of $\rho_{e}$ values considered
to the interval $[0,1/2]$ and we can also fix arbitrarily
the sign of $U$. We choose $U>0$, i.\ e.\ according to (\ref{ham})
the ions and electrons attract each other at the same lattice
site. Thus it is natural to refer to the system with
$\rho_{e}=\rho_{i}$ as {\em neutral}.

In this paper we study the properties of the Falicov-Kimball model
at zero temperature by means of an approximate method,
the {\em method of restricted phase diagrams}.
It is the set of the states available
for the system considered that is approximated (by a restricted
set of states), not the interactions.
We concentrate mainly on the {\em ion subsystem}, that is,
we  try to determine how the ions are
arranged in the ground state.
This sort of approach has already been used, first in \cite{jll},
then in \cite{frefal}, \cite{farbat}, \cite{guj}, \cite{llj},

The method of restricted phase diagrams is straightforward in
principle. In the {\em ca\-no\-ni\-cal ensemble},
we are interested in here,
some specified values are assigned to the densities $\rho_{e}$
and $\rho_{i}$ and the task amounts to minimizing
the set of functions
$\{e(w,\rho_{e}): w \in {\cal W}_{\rho_{i}} \}$
over the set ${\cal W}_{\rho_{i}}$ of
{\em admissible configurations}
that satisfy the condition $\rho_{i}(w)=\rho_{i}$.
The minimum function constitutes the required approximation to
the {\em ground state energy density} $e(\rho_{e},\rho_{i})$.
According to the thermodynamic stability conditions
$e(\rho_{e},\rho_{i})$ should be a convex function
of $(\rho_{e},\rho_{i})$.
Since the functions $\rho_{e} \to e(w,\rho_{e})$ are convex and
the minimum of a set of convex functions is not necessarily
convex, to guarantee the thermodynamic stability
it is necessary to construct the convex envelope
of the minimum function.
It is clear that the choice of ${\cal W}_{\rho_{i}}$
is decisive for the quality of the approximation.
We can choose ${\cal W}_{\rho_{i}}$ to be for instance
the set ${\cal W}_{\rho_{i},p}$
of all periodic ion configurations whose period does not exceed
some $p$ and $\rho_{i}(w)=\rho_{i}$. Then those points of the
graph of the minimum function that coincide with the graph of
the convex envelope can be associated with the periodic phases.
The remaining points can be  obtained as convex combinations
of some other points belonging to the graph of the minimum
function  and therefore they are associated with mixtures of periodic
phases. The ca\-no\-ni\-cal restricted phase diagrams of this kind
have been obtained in \cite{frefal} and \cite{farbat}.
It turns out however that the resulting ca\-no\-ni\-cal
restricted phase diagrams are highly unstable.
There are two main reasons for lack of stability in these
diagrams. First, they do not allow for mixtures of periodic
phases with different ion densities.
In many instances
the periodic phase corresponding to the point $(\rho_{e},\rho_{i})$
of the restricted phase diagram can be shown to have higher energy
than a suitable mixture of two or three periodic phases whose ion
densities are different from $\rho_{i}$
(cf. \cite{guj} and section 3).
Second, the sets of admissible ion configurations used there are
too restricted.

To overcome the stability problems mentioned and to avoid
dealing with mixtures we have studied the system in the grand
ca\-no\-ni\-cal ensemble and then translated the results into
the ca\-no\-ni\-cal ensemble.
In the {\em grand ca\-no\-ni\-cal ensemble} one is given the
electron and ion
chemical potentials $(\mu_{e}, \mu_{i})$ and the task is to find the
minimum function of the set
$\{f\left( w;\mu_{e}, \mu_{i} \right): w \in {\cal W} \}$
of the free energy densities
$f\left( w;\mu_{e}, \mu_{i}\right)$, where
\begin{equation}
\label{fen_w}
f\left( w;\mu_{e}, \mu_{i} \right)
=
e^{(U)}\left( w ; \mu_{e} \right) -
\mu_{e} \ \rho_{e}
\left( w; \mu_{e} \right)
- \mu_{i} \ \rho_{i}(w)
\end{equation}
and ${\cal W}$ is a set of admissible configurations.
The thermodynamic stability requires the minimum function to be
a concave function of $(\mu_{e},\mu_{i})$.
Since the free energy $f\left( w;\mu_{e}, \mu_{i} \right)$
is a concave function of $(\mu_{e},\mu_{i})$
and the minimum function of any set
of concave functions is concave, the thermodynamic stability
condition is satisfied automatically.
We have taken for the class ${\cal W}$
all periodic ion configurations whose periods do not exceed
16 lattice constants, typically, however we have also constructed
some diagrams where the maximal period has been increased to 40.

The ca\-no\-ni\-cal phase diagrams have a $U$-independent inversion
{\em symmetry} about the point $(1/2,1/2)$.
It follows from (\ref{symm_i}) and (\ref{symm_e}) that,
if $\stackrel{\sim}{w}$
is the ground state configuration of the ions at the point
$(\rho_{e},\rho_{i})$, then the adjoint configuration
$\stackrel{\sim}{w}^{*}$
is the ground state configuration at $(1- \rho_{e},1- \rho_{i})$.

%********************************************************************

\section{Some remarkable phases and their ions'
con\-fi\-gu\-ra\-tions}

To each point $(\rho_{e},\rho_{i})$ the canonical phase diagram
assigns a phase (more precisely a class of equivalent phases
that consists of the phases related by the symmetries of the
lattice), which is called here $(\rho_{e},\rho_{i})$-phase
or the {\em stable phase}.
Since we are concerned mainly with the ion subsystem, we
characterize the phases by their ions' arrangements $w$
and the electron density $\rho_{e}$.
{}From a few thousand of the periodic ion configurations that
have been tested, each time a restricted phase diagram was
constructed, only a few per cent, at most, happen to be the ions'
arrangements of the phases that appear in the phase diagram.
Many phases turn out to be the members of some infinite
families of phases whose periodic ion configurations comply with
peculiar rules and the ion and electron rational densities
satisfy linear relations.

There are of course only two translationally invariant ion
configurations, the {\em full configuration},
where all the sites of the lattice are occupied ($w(x)=1$)
and the {\em empty configuration} with no ions at all ($w(x)=0$).
The phases with translationally invariant ion configurations are
called the {\em full phases} and the {\em empty phases},
respectively.
The unique band of the empty configuration extends from $-2$ to
$2$,
while the band of the full configuration is obtained by
translating the latter by $-U$.
In the full or empty phases the system is a conductor, unless
$\rho_{e}=0$ or $1$.
The spectrum of any ion configuration is confined to the
interval given by the lower band edge of the full configuration,
$-2-U$, and the upper band edge of the empty configuration, 2.
For $U>4$ the two bands do not overlap and the gap separating them
opens in the spectrum of any other configuration.

{}From now on when we speak of periodic phases or periodic ion
configurations we mean that their period is at least 2.

The infinite families of the
{\em ion n-molecule most homogeneous phases} $[p/q]_{n}$
are characterized by the so called
{\em ion $n$-molecule most homogeneous configurations}
of ions and the relation  $\rho_{i}=n \rho_{e} $, where
$n$ is an integer, $\rho_{e}=p/q$  and $ 0 \leq \rho_{e} \leq 1/n$.
In the context of the Falicov-Kimball model
the ion $n$-molecule  most homogeneous configurations have
originally been defined in \cite{lem} (the case $n=1$)
and in \cite{guj} ($n \geq 2 $).
Let $p$ and $q$, $p<q$, be relatively prime integers.
The unit cell of the ion $n$-molecule
most homogeneous configuration,
whose ion density is $\rho_{i}=np/q$, has the length $q$ and
the arrangement of the ions is given by
the solutions $k_{j}$  to the equations
\be
\label{ip_mh}
p \, k_{j}=j \; mod \: q, \; \;j=0, 1, \ldots, np-1.
\ee
The ions form clusters, the ion $n$-molecules, that consist
of $n$ consecutive sites occupied by the ions.
The relation between the ion and electron densities means
that in these phases there is one electron per ion $n$-molecule.
The distances between any two nearest neighbour clusters are either
$d$ or $d+1$.
Similarly the distances between next nearest
neighbour $n$-molecules are either $d'$ or $d'+1$, and so on.
In \cite{guj} the ion $1$-molecule most homogeneous configurations
have been called the
{\em atomic most homogeneous configurations}.
The following example illustrates the definitions introduced:
for $p=2$ and $q=7$ the unit cell of  $[2/7]_{1}$ phase is
$[\bullet\circ\circ\circ\bullet\circ\circ]$,
the unit cell of $[2/7]_{2}$ phase is
$[\bullet\bullet\circ\circ\bullet\bullet\circ]$,
and the unit cell of $[2/7]_{3}$ phase is
$[\bullet\bullet\bullet\bullet\bullet\bullet\circ]$.

The infinite families
of the {\em hole n-molecule most homogeneous phases} $[p/q]_{n}^{*}$
are characterized by the
{\em hole $n$-molecule most homogeneous configurations}
of the ions and the relation $ 1- \rho_{i} =n \rho_{e} $, where
$n$ is an integer, $\rho_{e}=p/q$, $ 0 \leq \rho_{e} \leq 1/n$.
The hole $n$-molecule most homogeneous configuration
is defined in a similar way as the ion $n$-molecule most
homogeneous configuration,
except that the ions are replaced by the holes.
In this sense the
star can be interpreted as the operation of taking adjoint.
Clearly the adjoint of the adjoint phase is the original
phase.
In these phases there is also one electron per molecule but
this time it is the hole $n$-molecule.

In each family of the phases described so far
the corresponding periodic ion arrangements can be viewed
as composed of the same sort of molecules and there is one
electron per molecule.
There appear however in the phase diagrams other families of
phases, where the corresponding periodic ion arrangements
can be viewed as composed of the hole $n$-molecules,
with $n \geq 2$,
and of the composite molecules that consist of 1 ion, 1 hole and
1 electron.
In these phases there is also one electron
per each composite molecule but the number of electrons per hole
molecule depends on its size.
Specifically for a pair of
irreducible rationals $p/q$ and $h/k$, such that
$1/3 \leq h/k \leq p/q \leq 1/2$ and $k-2h=1$,
the $[p/q]_{h/k}$ phase has period $q$,
the electron density $\rho_{e}=p/q$,
the ion density $\rho_{i}=k \rho_{e} - h$,
and the positions of the ions in its unit cell are given by
the solutions $k_{j}$  to the equations
\be
\label{ip_c}
p\, k_{j}=j \; mod \: q,\; \; j=0, 1, \ldots, kp - hq - 1.
\ee
The rationals $h/k$ of $[p/q]_{h/k}$ phases that are stable in
our restricted diagrams belong to a special sequence, which will
be specified later.
The $[p/q]_{h/k}$ phase can be viewed as a periodic arrangement of
the hole $k$-molecules with $h$ electrons per hole molecule
and composite molecules built of 1 ion, 1 hole and $1$ electron.
This definition of $[p/q]_{h/k}$ families of phases does not apply
to the limiting case $p/q=h/k=1/2$ whose definition is postponed
to the text that follows.

In the phase diagrams there appear also the families of phases
that are adjoint to the $[p/q]_{h/k}$ phases.
Every $[p/q]_{h/k}$ phase has its adjoint phase $[p/q]_{h/k}^{*}$,
where the hole $k$-molecules are replaced by the ion $k$-molecules.
The ion and electron densities of the $[p/q]_{h/k}^{*}$ phase
satisfy the linear relation $ \rho_{i}= - k \rho_{e} + k - h $.

One of the remarkable features of the families of periodic phases
described so far is that their density pairs $(\rho_{e},\rho_{i})$
and their ions' configurations can be obtained recursively,
following a recursive construction of the so called
Farey sequence. Therefore we can speak of the
{\em Farey sequences of phases} like the Farey sequence of
$[p/q]_{n}$ phases or the Farey sequence of $[p/q]_{h/k}$ phases,
etc.

The {\em Farey sequence} ${\cal F}_{n}$ {\em of order} $n$
is the ascending sequence of irreducible rational fractions
between $0$ and $1$
whose denominators do not exceed $n$, i.e. the fraction $p/q$,
with $p$ and $q$ relatively prime integers, belongs to ${\cal F}_{n}$
if $0\leq p \leq q \leq n $ \cite{harwri}.
Having any two consecutive elements $p/q$ and $p'/q'$ of
${\cal F}_{n}$, the "parents",
we can form their "descendant" $p''/q''$,
\be
\label{vec_desc}
\frac{p''}{q''}= \frac{p + p'}{q + q'},
\ee
which satisfies the inequalities
\be
\frac{p}{q} < \frac{p''}{q''} < \frac{p'}{q'}.
\ee
This suggests how to construct ${\cal F}_{n}$ recursively.
To build up ${\cal F}_{n}$ we start with ${\cal F}_{1}$,
which is simply the pair of the "first parents" $\{ 0/1, 1/1 \}$,
and add to it the descendant $1/2$ of the "first parents",
what results in ${\cal F}_{2}$.
Then we choose a new pair of "parents" and repeat this step.
The process is continued until the denominators of new
"descendants" exceed $n$.
Let $p/q$ and $p'/q'$ ($p/q < p'/q'$)
be two consecutive elements of ${\cal F}_{n}$, for some $n$.
Replacing the pair of the "first parents" $\{ 0/1, 1/1 \}$
by the pair $\{p/q,p'/q'\}$ and continuing the recursive process
described until the denominators of new "descendants" exceed some
$n' > n$, we end up with
the {\em Farey sequence of order} $n'$
{\em generated by the "parents"} $\{ p/q,p'/q' \}$.
The sequence so defined consists of those elements $p''/q''$
of ${\cal F}_{n'}$
that satisfy the inequalities $p/q \leq p''/q'' \leq p'/q'$.
This definition can be extended to the vector case.
Namely starting with the pair of vectors
$\{ (p/q, r),(p'/q',r') \} $ (with $p/q, p'/q'$ as above),
where $ 0 \leq r,r' \leq 1 $ are rationals,
and forming "descendants" componentwise we arrive at
the {\em vector Farey sequence of order} $n'$ ($n'>n$)
{\em generated by the pair of vector "parents"}.
The recursion step is repeated
until the denominators of the first components of the
"descendants" exceed $n'$.

For each family of the phases described so far the density pairs
$(\rho_{e},\rho_{i})$ are of the form $(p_{e}/q_{e},p_{i}/q_{e})$,
where $p_{e}/q_{e}$ is an irreducible rational, and the
components of these density pairs satisfy
a linear relation $a\rho_{e}+b\rho_{i}=c$.
Clearly,
if $(\rho_{e}, \rho_{i})$ and $(\rho_{e}', \rho_{i}')$ are two
such pairs, then their "descendant" $(\rho_{e}'', \rho_{i}'')$
(defined componentwise) satisfies the same linear relation.
Therefore the density pairs of the families of phases considered
constitute the vector Farey sequences.
Specifically the density pairs of the $[p/q]_{n}$ phases
form the vector Farey sequence generated by
the "first parents" $\{ (0/1, 0/1), (1/n, n/n) \}$,
the density pairs of $[p/q]_{n}^{*}$ phases form the vector
Farey sequence generated by $\{ (0/1, 1/1), (1/n, 0/n) \}$,
while those of the $[p/q]_{h/k}$ and $[p/q]_{h/k}^{*}$ phases
form the vector Farey sequences generated by
$\{ (h/k, 0/k), (1/2, 1/2) \}$ and $\{ (h/k, k/k), (1/2, 1/2) \}$,
respectively, provided that $h/k$ and $1/2$ are "parents" in
${\cal F}_{k}$.

It appears that following these recursive procedures
one can construct also the unit cells of the phases considered.
For that it is enough to specify the  unit cells of
the "first parents" and to identify the operation of forming
descendants with the simple concatenating of these unit cells.
In the case of $[p/q]_{n}$ phases the unit cell of the $(0/1,0/1)$
phase is an empty site with no electrons $[\circ]$,
while the unit cell of the $(1/n,n/n)$ phase consists of
$n$ consecutive occupied sites with one
electron $[\overbrace{\bullet\bullet\ldots\bullet}^{n}]$.
Thus, for $n=2$
their descendant has the densities $(1/3, 2/3)$ and its unit cell
is $[\circ\bullet\bullet]$,
the descendant of $\{ (0/1, 0/1), (1/3, 2/3) \}$ has
the densities $(1/4,1/2)$ and the unit cell
$[\circ\circ\bullet\bullet]$ while the descendant of
$\{ (1/4,2/4), (1/3,2/3) \}$ has the densities $(2/7,4/7)$ and
the unit cell $[\circ\circ\bullet\bullet\circ\bullet\bullet]$, etc.
In the case of $[p/q]_{h/k}$ phases the unit cell of $(h/k, 0/k)$
phase consists of $k$ unoccupied sites with $h$ electrons and that
of $(1/2,1/2)$ phase consists of two sites with one ion and one
electron.

The $[p/q]_{h/k}$ phases seem to be natural generalizations
of the $n$-molecule most homogeneous phases. They are generated
just by different "first parents". Therefore it is tempting to
name the phases generated by some pair of "first parents" phases
by means of the recursive concatenating the
{\em generalized most homogeneous phases}.

If on the other hand, given a pair of "first parents" phases,
we first concatenate some number of the unit cells of one parent
and then we concatenate the resulting composite cell with a similarly
obtained composite cell of the other parent, we obtain a phase,
which, compared to the corresponding most homogeneous phase,
is extremely inhomogeneous.
We term it {\em the generalized crenel phase}.
The {\em crenel phases} considered in \cite{guj} are
generated by the unit cells $\{[\circ], [\bullet]\}$.
They are the "most inhomogeneous" phases among all the periodic
phases with given density and period.
It is clear that the families of the generalized most homogeneous
phases and the generalized crenel phases are not disjoint.

Now we are in a position to define another family of
the generalized most homogeneous phases, denoted $[p/q]_{1/2}$.
All the members of this family have the same electron density
$\rho_{e}=1/2$ and therefore they are characterized by their
ion density $\rho_{i}=p/q$.
Using the recursive procedure of forming descendants,
the density pairs of these phases can be obtained
from the "first parents" $\{(1/2,0/2),(1/2,1/2)\}$
while their unit cells are generated by the "first parents"
unit cells of the form
$[\circ\circ]$ with 1 electron and $[\bullet\circ]$ with 1 electron.
For instance the ion densities of the $[p/q]_{1/2}$ phases with
$1/2 \leq p/q \leq3/4$ and whose periods do not exceed 14
constitute the following sequence (in the ascending order):
$1/2$, $8/14$, $7/12$, $6/10$, $5/8$, $9/14$, $4/6$, $7/10$,
$10/14$, $3/4$. The denominators of these fractions are equal
to the periods of the corresponding phases. All these phases,
except those whose ion densities are $9/14$, $7/10$ and $10/14$,
are simultaneously of the generalized crenel type.

The only {\em aperiodic phases} that appear in our restricted phase
diagrams are those that constitute some mixtures of a few
periodic or translationally invariant phases.
The ion configurations of the mixtures are the corresponding mixtures
of the ion configurations of the component phases \cite{guj}.
A distinguished role is played by the mixtures of the translationally
invariant  phases,
the so called {\em segregated phases} \cite{frefal}.
For $\rho_{e}<1/2$ only the mixtures of the full phases
and the vacuum turn out to be stable \cite{guj}.
Their ion configurations are the corresponding mixtures
of the full and empty configurations,
i.\ e. all the ions clump together. It has been proved in
\cite{lem}
that if $\rho_{e} \neq \rho_{i}$ then there is a sufficiently
large $U$ such that above it the segregated phase is stable.
In \cite{guj} the exact stability region of the segregated phases
has been conjectured. This region consists of those points
$(\rho_{e},\rho_{i})$ that satisfy the inequality
$\rho_{e}<\rho_{i} b(U)$, where $b(U)$ is the solution to the
equation derived in \cite{guj}.
More details concerning these phases can be found in \cite{frefal},
\cite{guj}, \cite{brandt}, \cite{lem}.

%******************************************************************

\section{Main observations and conjectures}

The excerpts from our data are presented in Figs 1--7.
First, in Figs 1a and 1b, we show two schematic phase diagrams, which
indicate where the families of characteristic phases, described in
the previous section, are located in the $(\rho_{e},\rho_{i})$-plane.

Then in Figs 2--6 we show the actual data in the form of the
restricted canonical phase diagrams.
The set of the admissible ion configurations used to obtain these
diagrams consisted of all the periodic ion configurations with the
period not greater than 14 and of the full and empty configurations.
In these diagrams the $(\rho_{e},\rho_{i})$-phases are represented
by the black spots. In order to determine their rational $\rho_{e}$
and $\rho_{i}$ coordinates it is sufficient to know that
the $\rho_{e}$ values of the stable phases constitute the
following sequence of irreducible rationals
(in the ascending order):\\
$1/14$, $1/13$, $1/12$, $1/11$, $1/10$, $1/9$, $1/8$, $1/7$, $2/13$,
$1/6$, $2/11$, $1/5$, $3/14$, $2/9$, $3/13$, $1/4$, $3/11$, $2/7$,
$3/10$, $4/13$, $1/3$, $5/14$, $4/11$, $3/8$, $5/13$, $2/5$, $5/12$,
$3/7$, $4/9$, $5/11$, $6/13$, $1/2$.
Moreover, if $\rho_{e}=p_{e}/q_{e}$ is any element of this sequence,
then the $\rho_{i}$ coordinates of the majority of the black spots,
whose $\rho_{e}$ coordinates are equal to $p_{e}/q_{e}$,
belong to the sequence
$1/q_{e}$, $2/q_{e}$, \ldots, $(q_{e}-1)/q_{e}$.
For the set of the admissible ion configurations used to obtain
the diagrams presented the exceptions are rare (most of them are
the phases with $\rho_{e}=1/2$) and they are indicated in the
captions of Figs 2--6.
The periodic ion configurations corresponding to the
$(\rho_{e},\rho_{i})$-phases with the $\rho_{e}$ and $\rho_{i}$
coordinates belonging to the sequences quoted can easily be
determined.
Suppose that $\rho_{e}=p_{e}/q_{e}$ and $\rho_{i}=p/q_{e}$.
Then the unit cell has the length $q_{e}$.
If we label the positions of the lattice sites in the
unit cell $0$, $1$, \ldots, $p - 1$, then the positions of the ions
are $0$, $k_{1}$, \ldots, $k_{p-1}$, where $k_{1}$ is the
inverse of $p_{e}$ mod $q_{e}$ and $k_{i}= i k_{1}$ mod $q_{e}$
(cf the equations \ref{opti2} and the comments of point 2
in the sequel).

Finally in Fig. 7 we show the
intersections of the restricted canonical phase diagrams with the
line $\rho_{i}=1/2$, obtained for different $U$ values.

{\bf 1. } The canonical phase diagram establishes a one to one
$U$-dependent correspondence between
the density pairs $(\rho_{e},\rho_{i})$ and the stable phases.
Suppose that at the point $(\rho_{e},\rho_{i})$,
where $\rho_{e}=p_{e}/q_{e}$ and $\rho_{i}=p_{i}/q_{i}$
are irreducible rationals, the stable phase is periodic.
The stable phase has necessarily the lowest Fermi sea energy.
Now according to the second order degenerate perturbation theory
with respect to $U$, for given  $\rho_{e}=p_{e}/q_{e}$
the lowest Fermi sea energy among the periodic ion arrangements
with $\rho_{i}=p_{i}/q_{i}$
is attained by the periodic configuration that has a gap
at the Fermi level and this gap is the largest \cite{ashmer}.
Specifically, to meet the requirement of the maximal gap,
the period of the phase corresponding
to $(p_{e}/q_{e},p_{i}/q_{i})$ has to be equal to
$Q$ -- the least common multiple of
$q_{e}$ and $q_{i}$ and the configuration $w$ has to maximize
$|\stackrel{\wedge}{w}(2\pi \rho_{e})|$,
where
\be
\label{opti1}
\stackrel{\wedge}{w}\left(2\pi \frac{p_{e}}{q_{e}}\right) =
\sum_{x=0}^{Q-1} \exp\left(i2\pi \frac{p_{e}}{q_{e}} x \right)\,
w(x).
\ee
The same conclusion has been reached in \cite{frefal}.
If there  is only one phase that complies with the rules quoted,
we call it {\em uniform} (this is always the case
if $Q=q_{e}$),
if there is more than one, we call them
{\em degenerate uniform}.
For all $U$ values tested, from $0.2$ up to $6$,
our restricted phase diagrams conform to these rules
surprisingly well.

First, for all values of $U$, $0.2 \leq U \leq 6$,
all the phases have appeared to be insulators.
However, in order to reach this situation for small values of $U$
the maximal period of the configurations tested has to be quite
large.
For instance for $U=0.2$, the $(1/2,1/2)$-phase is stable also
for $\rho_{e}$ in a small vicinity of $1/2$,
unless we take into account in the vicinty of $(1/2,1/2)$
the phases whose periods go up to about forty.
Since in the ca\-no\-ni\-cal phase
diagrams presented in \cite{frefal} the maximal period of the
configurations tested is only $8$, for small $U$ the phases
that appear are not insulators. In the diagrams presented in
\cite{farbat} this effect is even more pronounced, occurs also
for $U$ much larger than 1. This is a consequence of using too
"directed" set of admissible ion configurations.

Second, only for intermediate $U$ values, $1< U < 3.6$ roughly,
we have detected some {\em nonuniform} phases.
One example is the $(3/7,1/2)$-phase
whose ion configuration is not among those for which
the ma\-xi\-mum of $|\stackrel{\wedge}{w}(2\pi \rho_{e})|$ is
attained.
It is visible for instance
in $U=3$ phase diagram, Fig.6.
Its period 14 is the least common multiple
of $q_{e}$ and $q_{i}$, however its unit cell is of the form
$[\bullet\circ\circ\bullet\circ\bullet\circ
\circ\bullet\circ\bullet\bullet\bullet\circ]$,
while the $(3/7,1/2)$ phase that maximizes
$|\stackrel{\wedge}{w}(2\pi \rho_{e})|$
has the unit cell
$[\bullet\circ\circ\bullet\circ\bullet\circ
\bullet\circ\bullet\bullet\circ\bullet\circ]$.
Other examples of nonuniform phases have been revealed by
additional studies, not represented in the diagrams included.
For $U$ between $1$ and $1.9$ there appear
$(1/2,\rho_{i})$-phases whose ion configurations are of the
generalized crenel type,
generated by the "parents" : $[\circ\bullet]$ with
one electron and $[\bullet\bullet]$ (or $[\circ\circ]$)
with one electron.
The periods of these phases are not the least common multipliers
of $q_{e}$ and $q_{i}$.\\

{\bf 2. } For every $\rho_{e}=p_{e}/q_{e}$ there is
a characteristic $U$ value, $U_{1}(p_{e}/q_{e})$,
such that below $U_{1}(p_{e}/q_{e})$
the general rule of point 1
is accompanied by an additional one.
Namely, only the $(p_{e}/q_{e},p_{i}/q_{i})$-phases
such that  $Q=q_{e}$ appear in the restricted phase diagrams.
The smallest $U_{1}(p_{e}/q_{e})$ is $U_{1}(1/2)\simeq 1$.
As a consequence, below $U_{1}(p_{e}/q_{e})$
the positions of the ions in the unit cells of
$(p_{e}/q_{e}, p_{i}/q_{i})$-phases
are given by the solutions $k_{j}$  to the equations
\be
\label{opti2}
p_{e} \, k_{j}=j \; mod \: q_{e}, \; \; j=0, 1, \ldots,
p_{i} q_{e}/q_{i} - 1.
\ee
It is easy to see that once the $k_{1}$ solution has been found,
the remaining solutions are given by $k_{j}=j k_{1}$ mod $q_{e}$.
This additional rule turns out  to be satisfied also for
sufficiently large $U$ (above $U=3.6$ roughly),
independently of $\rho_{e}$,
simply because at these $U$ values the ion
arrangements of the periodic phases are the ion $1$-molecule
most homogeneous configurations.

Moreover, there is another characteristic $U$ value,
$U_{2}(p_{e}/q_{e})$, $U_{2}(p_{e}/q_{e}) \leq U_{1}(p_{e}/q_{e})$,
such that below $U_{2}(p_{e}/q_{e})$,
if $p_{i,min}/q_{e}$ is the minimal ion density
of the stable phases with $\rho_{e}=p_{e}/q_{e}$
and $p_{i,max}/q_{e}$ is the maximal one,
then all the remaining $(p_{e}/q_{e}, p_{i}/q_{e})$  phases
with $p_{i}$ assuming all the values
$p_{i,min}$, $p_{i,min}+1$, \ldots, $p_{i,max}$
are stable as well.
This is a more precise form of the observation made in
\cite{guj}. The value of $U_{2}(p_{e}/q_{e})$ decreases
as the period $q_{e}$ of the phases increases.
It is not clear if there is some nonzero $U_{2}$ that is good
for all periods.
In Fig.3 ($U=1.4$), for $\rho_{e}=5/14$ there appear the phases
corresponding to $p_{i}=5,6,9,10$ but those with $p_{i}=7,8$ are
missing. Similarly, for $\rho_{e}=4/11$ there appear the phases
corresponding to $p_{i}=4,5,7,8$ but that with $p_{i}=6$ is
missing. Some other examples are visible in this diagram.
All these missing phases are stable already in the phase diagram
for $U=0.8$, Fig.2.
The values of $p_{i,min}$ and $p_{i,max}$ appear to be determined
by the $n$-molecule most homogeneous phases and the generalized
most homogeneous phases.
For given $\rho_{e}= p_{e}/q_{e}$ the minimal
$p_{i,min}/q_{e}$ and the maximal $p_{i,max}/q_{e}$ ion densities
are determined by the minimal and maximal $\rho_{i}$ values
of the intersections of the line $\rho_{e}= p_{e}/q_{e}$ and the
line segments formed by the $[p/q]_{n}$, $[p/q]_{n}^{*}$,
$[p/q]_{h/k}$ and $[p/q]_{h/k}^{*}$  stable phases.

{\bf 3. } When the maximal period of the admissible ion
configurations grows, then the number of the stable phases in the
Farey sequences of the most homogeneous and the generalized
most homogeneous phases grows as well.
It appears that the order of any such a Farey sequence of phases
is equal to the maximal period of the admissible ion configurations.
Moreover the Farey sequences of the most homogeneous and
the generalized most homogeneous phases are stable in specific
$U$-dependent intervals of $\rho_{e}$ values.
Only the $\rho_{e}$ intervals of $[p/q]_{1}$ phases
are nonvoid for all $U$.
In the  case of the ion $n$-molecule
most homogeneous phases with $n>1$ the critical $U$ values at which
the $\rho_{e}$ intervals open have been determined in \cite{guj}.
The hole $n$-molecule most homogeneous phases appear in
the phase diagrams together with
the $2$-molecule most homogeneous phases, i.e.  (according to
\cite{guj}) below $U=2.66$.
The $[p/q]_{n}$ families appear in the phase diagrams in succsession,
starting with $n=2$, so that for sufficiently small $U$
only the phases with $n$ not exceeding some characteristic value
are stable \cite{guj}. In contrast, for given $U< 2.66$
all the $[p/q]_{n}^{*}$ phases with $n$ greater than some
characteristic value appear. For instance for $U=2$ (Fig.4),
those with $n=1$ are still missing, but somewhere between $U=2$
and $U=1.4$ all the $[p/q]_{n}^{*}$ families become stable.
The $\rho_{e}$ stability intervals of the $[p/q]^{*}_{n}$ family
of phases are determined by the points of intersection
of the line $\rho_{i}=1 - n\rho_{e}$  with the segments of
those lines $\rho_{i}=n'\rho_{e}$ that correspond to the stable
phases.

{\bf 4. } The Farey sequences of the $[p/q]_{h/k}$
generalized most homogeneous phases
and their adjoint phases start to appear somewhere between $U=1$
and $U=0.8$, as $U$ decreases.
In Fig.2 ($U=0.8$) three representatives of the $[p/q]_{1/3}$
phases and also three representatives of the adjoint family
are visible. For $U=0.4$ two families, $[p/q]_{1/3}$  and
$[p/q]_{2/5}$ and their adjoints are well represented in our
restricted phase diagrams
and for $U=0.2$ a few representatives of
the $[p/q]_{3/7}$ family appear.

{\bf 5. } For $U$ between $1$ and $3.6$ we have found a family
of phases with $\rho_{e}=1/2$.
For $U>2$ this family consists of the generalized most homogeneous
phases of the $[p/q]_{1/2}$ type (and the adjoint family).
For instance, in Fig.6 ($U=3$) and in Fig.5 ($U=2.6$) this family
is represented by the phases whose ion densities
(in the ascending order) are $8/14$, $7/12$, $6/10$, $5/8$, $9/14$
and $4/6$,
where the denominators are equal to the periods of these phases.
The phases that correspond to all $\rho_{i}$ fractions quoted,
except $9/14$ are uniform (the numerators of these $\rho_{i}$
fractions exceed the halves of the denominators by one)
and their unit cells are of the generalized crenel type.
In contrast the $(1/2,9/14)$-phase is degenerate uniform
and it is not generalized crenel. Its unit cell has the form
$[\circ\bullet\circ\bullet\bullet\bullet
\circ\bullet\circ\bullet\circ\bullet\bullet\bullet]$,
in agreement with the recursive concatenating procedure.
For sufficiently small $U$ the degenerate uniform $[p/q]_{1/2}$
phases that are not generalized crenel transform into
(degenerate uniform) phases of the generalized crenel type
(with the same ion density).
Thus in some points $(1/2,\rho_{i})$ we observe
the {\em structural phase transitions} --
one stable periodic phase transforms into another stable periodic
phase as $U$ varies.
For instance, for $U=2.1$ the  $(1/2,9/14)$-phase has the unit cell
shown above. Already for $U=1.9$ the unit cell of the
$(1/2,9/14)$-phase is of the generalized crenel type:
$[\circ\bullet\circ\bullet\circ\bullet\circ\bullet\circ\bullet
\bullet\bullet\bullet\bullet]$.
Another manifestations of the effect described can be found in the
diagrams for $U=2$ (Fig.4) and for $U=1.4$ (Fig.3).
In Fig.4 the $(1/2,\rho_{i})$-phase with the largest ion density,
$\rho_{i}=7/10$, is degenerate uniform and generalized crenel.
Its unit cell has the form
$[\circ\bullet\circ\bullet\circ\bullet\bullet\bullet\bullet\bullet]$,
while the unit cell of the generalized most homogeneous phase
$[7/10]_{1/2}$ is
$[\circ\bullet\circ\bullet\bullet\bullet\circ\bullet\bullet\bullet]$.
The remaining $(1/2,\rho_{i})$-phases, with $\rho_{i}$ equal to
$5/8$, $6/10$, $7/12$ and $8/14$,
are uniform and simultaneously generalized most homogeneous and
generalized crenel.
The only $(1/2,\rho_{i})$-phase visible in Fig.3,
with $\rho_{i}=10/14$, is also degenerate uniform and generalized
crenel.
Its unit cell is
$[\circ\bullet\circ\bullet\circ\bullet\circ
\bullet\bullet\bullet\bullet\bullet\bullet]$,
while the $[10/14]_{1/2}$-phase has the unit cell
$[\circ\bullet\circ\bullet\circ\bullet\bullet\bullet\bullet\bullet
\circ\bullet\bullet\bullet]$.\\

{\bf 6. } When $U$ increases towards $4$, all the periodic phases
different from the $1$-molecule most homogeneous ones start to
dissappear (their domains in the $(\mu_{e},\mu_{i})$ plane shrink
quickly).
Thus, our diagrams suggest that above $U>4$ the only phases
that remain in the phase diagrams are
the ion $1$-molecule most homogeneous phases with arbitrary
$\rho_{e}$,
the mixtures of the ion $1$-molecule most homogeneous phases
and the full phases or the empty phases, and the segregated phases.
This is in contrast with the ca\-no\-ni\-cal phase diagrams of
\cite{frefal} and \cite{farbat}.\\

{\bf 7. } The "evolution" of the phase diagrams with decreasing
$U$ can be viewed as building up an extra symmetry -- the ion
(or the electron) hole-particle symmetry.
In the $(\rho_{e},\rho_{i})$-plane a polygon is formed,
which is invariant with respect to the reflections in
the line $\rho_{i}=1/2$.
The segments of the lines of the ion and hole most
homogeneous phases determined by the mutual intersections
of these lines constitute the sides of that polygon.
For instance, in the $U=1.4$
restricted phase diagram (Fig.3) the tetragon with the following
vertices $(1/2,1/2)$ $(1/3,2/3)$ $(1/4,2/4)$ is well visible,
however it is not yet reflection invariant, since some phases inside
this tetragon are missing, but for $U=0.8$ (Fig.2) the reflection
symmetry with respect to the line $\rho_{i}=1/2$ is already restored.
As $U$ decreases the invariant polygon gradually extends through out
the smaller $\rho_{e}$ values.
For $U=0.4$ it consists of two tetragons whose boundaries are visible
in the schematic phase diagram, Fig.1.
The appearance of the hole most homogeneous phases $[p/q]^{*}_{n}$,
the generalized most homogeneous phases $[p/q]_{h/k}$ and
$[p/q]^{*}_{h/k}$ phases, as well as
the fact that, as $U$ decreases the minimal $\rho_{e}$ values
for which the $[p/q]_{n}$ phases are stable increase, can be viewed
as manifestations of the tendency to restore the ion hole-particle
symmetry. Because of the inversion symmetry of the phase diagram,
the ion hole-particle symmetry is equivalent to the electron
hole-particle symmetry.\\

{\bf 8. } In agreement with the Gibbs phase rule for a two
component system at a fixed temperature, only mixtures of two
or three phases appear in the restricted ca\-no\-ni\-cal
phase diagrams.
Actually in the diagrams presented, Figs 2--6, there appear also
mixtures of more than three periodic phases.
This is a consequence of the finite step with which
the electron chemical potential axis has been scanned, when the grand
canonical phase diagrams were constructed.
The stable mixtures can be arranged in five groups.
For $\rho_{e}<1/2$ there appear mixtures of:
(a) two periodic phases,  (b) three periodic phases,
(c) a periodic phase and a translationally invariant phase
(where the electron density of the latter may vary in an interval),
(d) two periodic phases with a translationally invariant phase
(whose electron density is uniquely determined by the periodic
phases) and
(e) the segregated phases that are mixtures of the vacuum
(i.e. the empty phase with no electrons) and the full
phases with $\rho_{e}$ not greater than certain $U$-dependent
critical value \cite {guj}, \cite{brandt}, \cite{lem}.
The mixtures belonging to the groups (a) and (b) are the insulators
while those belonging to the other groups are the conductors.
Typically the periodic phases that appear in the mixtures
of the kind (c) and (d) belong to the Farey sequences of the
(generalized) most homogeneous phases.
When the maximal period of the admissible ion configurations grows,
then the areas occupied by the mixtures of the kind (b) and (d)
seem to shrink quickly.
The mixtures of the kind (d), where the electrons densities of the
two periodic phases are equal and assume a limiting value of the
$\rho_{e}$ stability interval of one of the Farey sequences of
phases, seem to be the only exceptions. In Fig. 2 these are for
instance the mixtures of $(1/6, 1/3)$, $(1/6,1/2)$ and the full
phases, the mixtures of $(2/5,3/5)$, $(2/5,4/5)$ and the full
phases, etc.
Therefore we expect that in the limit of the infinite maximal period
there are no mixtures of the kind (b) and those of the kind (d)
are relatively rare.

{\bf 9. }  As $U$ increases, then typically
the periodic phases and their mixtures that are located at the
boundary of the $(\rho_{e},\rho_{i})$ region,
where the periodic phases and their mixtures are stable,
lose against mixtures of other periodic phases with a full or
an empty phase.
This results in an {\em insulator--conductor transitions}.
In the limit $U \to \infty$
the neutral phases (with arbitray $\rho_{e}$) remain the only
insulators.
However the insulator--conductor transitions can occur also
when $U$ decreases.
As first observed in \cite{guj}, as $U$ decreases
the $n$-molecular most homogeneous phases,
with electron densities sufficiently close to $0$ (or $1$)
lose their stability against mixtures of
$n'$-molecular most homogeneous phases ($n'>n$) with the empty
phases. In the case $n=1$ this occurs below $U=2/\sqrt{3}$
\cite{guj}.
This effect is visible for instance in Fig.2 ($U=0.8$).
In the region of small $\rho_{e}$ the
representatives of the neutral phases terminate for larger
$\rho_{e}$ than this is the case for the ion $n$-molecular phases
with $n=2,3$.

{\bf 10. }  If we start with the point $(p_{e}/q_{e},0)$
and move it up along the line $\rho_{e}=const$,
then typically it first passes through infinitely many
mixtures of some empty phases (with smaller $\rho_{e}$) and
some periodic ones (with larger $\rho_{e}$).
After that the point enters an interval of $\rho_{i}$ values,
where only periodic phases and their mixtures are the stable phases.
It is not clear if this interval contains only not more than
$q_{e}-1$ periodic phases with the period $q_{e}$
and mixtures of two consecutive periodic phases of this kind.
It may not be connected, there can be subintervals where mixtures
of some periodic phases and full or empty phases are stable.
Finally the point passes through infinitely
many mixtures of some periodic phases (with smaller $\rho_{e}$)
and some full phases (with larger $\rho_{e}$) until
it reaches eventually the full phase
(if $\rho_{e}>b(U)$,\cite{guj})
or the segregated phases (if $\rho_{e}<b(U)$).
Similar scenario,
but obtained by moving the point $(0,1/2)$ along the line
$\rho_{i}=1/2$ and then repeating this procedure for many values
of $U$, is presented in Fig. 7.
This kind of the restricted canonical phase diagram can directly
be compared with the diagrams shown in \cite{frefal}
and \cite{farbat}.\\

\section*{Acknowledgments}
The authors acknowledge the financial support of the
State Committee for Scientific Research (Poland) under grant
2 P302 147 04.
We thank Christian Gruber and Daniel Ueltschi for interesting
discussions.

\newpage

\centerline{ \bf Figure Captions}
\begin{description}
\item[{Fig. 1}] {
a)Schematic canonical phase diagram for $U=0.4$.
The shaded regions correspond to the segregated phases. The ion
$n$-molecule most homogeneous phases are distributed along
the continuous line segments while the hole $n$-molecule most
homogeneous phases along the dashed line segments.
The  $[p/q]_{h/k}$ phases are lying along the
dashed-doted line segments.\\
b)Schematic canonical phase diagram for $U=4$.
The only  periodic phases that appear in the diagram are
the ion $1$-molecule most homogeneous phases.
The points representing these phases constitute the line
segment that extends from $(0,0)$ to $(1,1)$.
}

\item[{Fig. 2}] {
The restricted canonical phase diagram for $U=0.8$.
This is a part of the complete diagram, which is sufficient for
reconstructing the remaining part by means of the inversion symmetry
(cf section 1).
The black spots represent the periodic phases
(whose period is at least 2).
Those black spots whose corresponding phases coexist in the grand
ca\-no\-ni\-cal phase diagram are joined by the straight line
segments. The points lying on such a segment represent the mixtures
of the two periodic phases that correspond to the ends of the
segment.
The straight line segments partition the phase diagram into convex
areas, most of which are triangles. These triangles correspond to
the coexistence points of three periodic phases in the grand
canonical phase diagrams.
The points located inside these triangles represent mixtures of the
three phases that correspond to the vertices of the triangles.
Some segments are missing and therefore there are areas which
according to this restricted diagram are mixtures of more than three
phases. This is an artefact related with the precision of the
diagram.
All the $(\rho_{e},\rho_{i})$-phases conform to the equations
\ref{opti2}.
}

\item[{Fig. 3}] {
The restricted canonical phase diagram for $U=1.4$.
This is a part of the complete diagram, which is sufficient for
reconstructing the remaining part by means of the inversion symmetry.
Only the $(1/2,10/14)$-phase does not conform to the equations
\ref{opti2}. It does not belong also to the $[p/q]_{1/2}$ family.
It is generalized crenel and its unit cell is
$[\circ\bullet\circ\bullet\circ\bullet\circ
\bullet\bullet\bullet\bullet\bullet\bullet]$.
See the caption of Fig. 2 and section 3 for more explanations.
}

\item[{Fig. 4}] {
The restricted canonical phase diagram for $U=2$.
This is a part of the complete diagram, which is sufficient for
reconstructing the remaining part by means of the inversion symmetry.
All the phases with $\rho_{e}=1/2$, except that with the largest
ion density $\rho_{i}=7/10$, belong to the family of
$[p/q]_{1/2}$ phases. Their ion densities (in the ascending
order) are: $8/14$, $7/12$, $6/10$, $5/8$.
The unit cell of the $(1/2,7/10)$-phase is
$[\circ\bullet\circ\bullet\circ\bullet\bullet\bullet\bullet\bullet]$.
See the caption of Fig. 2 and section 3 for more explanations.
}

\item[{Fig. 5}] {
The restricted canonical phase diagram for $U=2.6$.
See the caption of Fig. 2 and section 3 for more explanations.
All the phases with $\rho_{e}=1/2$ belong to the family of
$[p/q]_{1/2}$ phases. Their ion densities (in the ascending
order) are: $8/14$, $7/12$, $6/10$, $5/8$, $9/14$, $4/6$.
}

\item[{Fig. 6}] {
The restricted canonical phase diagram for $U=3$.
See the caption of Fig. 2 and section 3 for more explanations.
All the phases with $\rho_{e}=1/2$ belong to the family of
$[p/q]_{1/2}$ phases. Their ion densities (in the ascending
order) are: $8/14$, $7/12$, $6/10$, $5/8$, $9/14$, $4/6$.
The $(3/7,7/14)$-phase is not uniform.
}

\item[{Fig. 7}] {This is a schematic phase diagram,
the intersections of the restricted canonical phase diagrams
with the line $\rho_{i}=1/2$, obtained for the specific $U$ values
marked on the horizontal axis.
We can distinguish a few kinds of stable phases.
In the white region marked "segregated", the segregated phases are
the only stable phases.
In the light grey region marked "most homogeneous\&full"
a stable phase is a mixture of one or two
ion $n$-molecule most homogeneous phases and a full phase.
In the dark grey region marked "periodic phases and their
mixtures" the stable phases are either periodic or mixtures of one or
two periodic phases.
Finally in the unmarked grey region the stable phases
can be either like in the dark or light grey regions, or something
else, depending on $\rho_{e}$.
For the specific $U$ values marked on the horizontal axis, the
critical $\rho_{e}$ values, where the kind of stable phases changes,
are marked by the short horizontal segments. The continuous segments
correspond to periodic phases while the dotted ones to mixtures.
}
\end{description}

\end{document}